\documentclass[twocolumn,preprintnumbers,amsmath,amssymb]{revtex4}

\usepackage{dcolumn,epsfig}
\usepackage{bm}

\begin{document}
\title{Deterministic characterization of stochastic genetic circuits}

\author{Matthew Scott}\email{mscott@ctbp.ucsd.edu}
\author{Terence Hwa}\email{hwa@ucsd.edu}
\affiliation{
Center for Theoretical Biological Physics, Department of Physics, University of California, San 
Diego, \\
La Jolla, California, USA 92093-0374 
} %
\author{Brian Ingalls}
\email{bingalls@math.uwaterloo.ca}
\affiliation{
Department of Applied Mathematics, University of Waterloo, \\
Waterloo, Ontario, Canada N2L 3G1 
} %

\begin{abstract}
For cellular biochemical reaction systems where the numbers of molecules is small, significant noise is associated with chemical 
reaction events. This molecular noise can give rise to behavior that is very different from the predictions of deterministic rate 
equation models. Unfortunately, there are few analytic methods for examining the qualitative behavior of stochastic systems. Here 
we describe such a method that extends deterministic analysis to include leading-order corrections due to the molecular noise. The 
method allows the steady-state behavior of the stochastic model to be easily computed, facilitates the mapping of stability phase 
diagrams that include stochastic effects and reveals how model parameters affect noise susceptibility, in a manner not accessible 
to numerical simulation. By way of illustration we consider two genetic circuits: a bistable positive-feedback loop and a 
negative-feedback oscillator. We find in the positive feedback circuit that translational activation leads to a far more stable 
system than transcriptional control. Conversely, in a negative-feedback loop triggered by a positive-feedback switch, the 
stochasticity of transcriptional control is harnessed to generate reproducible oscillations.  \end{abstract}

\maketitle

Parallel advances in the conceptual understanding of gene regulation along with technological advances in molecular biology have 
given rise to the possibility of system-level quantitative kinetic measurements of living organisms~\cite{Amos} and synthetic 
genetic circuit designs~\cite{SynthHasty,SynthKaernBlakeCollinsReview}. Interpretation of time-series data from complex networks 
and reliable forward-design of gene circuits depend upon detailed quantitative mathematical models~\cite{SynthHasty,BintuMODELS}. 
These models generally take one of two largely exclusive forms -- either deterministic formulations with reactant concentration 
varying continuously in time and governed by a system of rate equations, or stochastic formulations that explicitly include the 
discrete and probabilistic change in reactant molecule numbers as each subsequent reaction occurs~\cite{Gillespie}. Both 
approaches have benefits and associated limitations. 

The great practical advantage of rate equation models is the ease with which the qualitative behavior of the system can be 
extracted. By focusing upon the long-term behavior, the model dynamics are simplified and one is able to gain insight into the 
expected response of the system~\cite{StrogatzBOOK}. Rate equation models, however, neglect the fact that chemical reaction 
networks are composed of species that evolve on discrete space -- jumping from some number of molecules to another as each 
reaction occurs~\cite{Kaern}. The resulting deviation from the deterministic formulation is called the \emph{intrinsic noise} in 
the system (since the fluctuations arise from the reaction dynamics themselves and not from some external source) 
~\cite{SwainSiggia, vanKampenBOOK}. In cellular systems with small numbers of reactant molecules, the relative magnitude of the 
intrinsic noise can be large, and can give rise to  
\emph{qualitatively} different behavior than what rate equation models would predict. A system that has several possible stable 
states, for example, may be induced to spontaneous transitions between them as a result of intrinsic noise~\cite{Sneppen,Wolynes}, 
leading to a stochastic switching of states.  In an excitable system, noise may cause oscillations to occur in a model that is 
otherwise stable~\cite{Vilar,Constructive,Elowitz2}. With a given set of physical parameters, it is possible to simulate 
explicitly the individual chemical reaction events, including the effect of intrinsic noise~\cite{Gillespie}. Nevertheless, the 
design of synthetic circuits, or therapeutics aimed at altering an existing network, require knowledge of the \emph{phase 
diagram}, which involves a systematic mapping of the parameter space. There, stochastic simulation becomes prohibitively 
time-consuming even for reasonably simple genetic circuits involving 2-3 genes (see below), and analytical methods are needed. 

 A number of analytic studies have been done recently to model intrinsic noise in genetic circuits, much of it built upon the 
linear noise approximation~\cite{vanKampenME} and focused upon the noise property itself, e.g., `noise propagation' through 
genetic networks~\cite{tenWolde,Pedraza}, the equilibrium distribution of fluctuations about multiple steady-states~\cite{Tomioka} 
and constructive effects of noise in signal processing~\cite{Paulsson,Constructive}. There has been comparatively little work, 
however, aimed at providing tools to study the effect of intrinsic noise on the stability of systems where stochastic models 
exhibit qualitatively different behavior from their deterministic counterparts~\cite{Deville}. Under these conditions, the linear 
noise approximation alone cannot predict qualitative changes in the observable dynamics of the system, as for example in the case 
of noise-induced oscillations~\cite{Elf}. Here we present an analytic method, which we call the \emph{effective stability 
approximation} (ESA), that extends the applicability of existing deterministic methods to include stochastic effects. The method 
is an extension of the linear noise approximation, including correction of stochasticity to the deterministic equations to the 
order $1/N$ (where $N$ is the number of molecules in the system). It conveniently connects deterministic and stochastic 
descriptions, allowing systematic exploration of parameter space while at the same time including the essential effect of 
intrinsic fluctuations. For the two model systems examined here, we find the ESA to capture reliably the essential features of 
those systems, correctly estimating the effect of intrinsic noise on the phase diagrams of systems dominated by as little as a few 
dozen molecules. 

ESA can be applied to generic models of genetic circuits, and a brief tutorial is presented in the {\it Methods} section with the 
hope that the approach can be used by other investigators to include stochastic effects in deterministic models. The full 
mathematical details are presented in the {\it Supplementary Material}. We illustrate the power of the method below by considering  
two examples - an autoregulator with positive feedback (an \emph{autoactivator})~\cite{Autoactivator} and an excitable genetic 
oscillator linking positive and negative feedback loops~\cite{Vilar,Atkinson}. The behavior of both circuits is conveniently 
visualized by means of a phase diagram that cannot be practically constructed using numerical simulations if stochastic effects 
are to be included. Furthermore, the analysis reveals that the system behavior is {\em completely} governed by a few dimensionless 
combinations of model parameters -- combinations that would be very difficult to infer from simulation data alone. We hope that 
our presentation of the ESA method will make it accessible to modelers, bioengineers and synthetic circuit designers for the 
analysis of various molecular circuits, while our description of the behaviors of the two model systems will provide 
quantitative-minded biologists with a concrete sense of the effect of stochasticity as well as a succinct means of 
characterization (e.g., a phase diagram with reduced variables). 

\section{Results \& Discussion}

\begin{figure}
\begin{center}
\epsfig{file=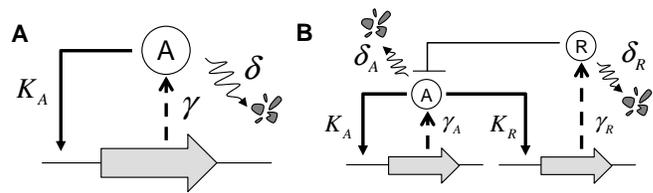,width=\linewidth}
\caption{(A) A positive-feedback loop capable of maintaining two stable states~\cite{Autoactivator}. (B) An excitable oscillator 
that exhibits noise-induced oscillations~\cite{Vilar,Atkinson}. The autoactivator triggers the production of a repressor $R$ that 
provides negative feedback control. (The dashed arrows denote lumped transcription and translation, the bold solid arrows denote 
activation, the blunt arrow denotes repression and the wavy arrows denote degradation.)}
\label{figureOne}
\end{center}
\end{figure}

\subsection{Autoactivator}

Perhaps the simplest circuit motif able to exhibit multiple stable states is the autoactivating positive feedback loop (Figure 
1a)~\cite{FerrellReview}. The circuit consists of a single gene encoding an activator. Several autoactivator circuits have been 
experimentally characterized, including the autoactivation of CI protein by the $P_{RM}$ promoter of phage $\lambda$ studied by 
Isaacs {\it et al.}~\cite{Autoactivator}, and the autoactivation of NtrC by the glnAp promoter of {\it E. coli} studied by 
Atkinson {\it et al.}~\cite{Atkinson}. The autoactivator circuit is expected to exhibit either a {\it HIGH} state characterized by 
an elevated level of protein synthesis, or a {\it LOW} state characterized by a low basal level of production. We simplify the 
model by assuming that the activator binding and mRNA turnover are fast compared to the lifetime of the protein activator. The 
effect of the activator is quantified by the \emph{activation function} $g\left({A/K_A,f}\right)$ where $A$ is the activator 
concentration, $K_A$ is the equilibrium dissociation constant of the activator and its cognate binding site, and $f$ is the 
maximum fold-activation in the circuit. As a particular example, we assume a Hill-form for the activation function 
$g\left({A/K_A,f}\right)$,
\begin{gather}
g\left( {\frac{A}
{{K_A }},f} \right) = \frac{{f^{ - 1}  + \left( {\frac{A}
{{K_A }}} \right)^n }}
{{1 + \left( {\frac{A}
{{K_A }}} \right)^n }},
\label{eq:Macro}
\end{gather}
with cooperative activation ($n=2$)~\cite{BintuMODELS}. The resulting model is a single kinetic equation governing the activator 
concentration $A(t)$~\cite{Keller,Autoactivator}(Figure 1a),
\begin{gather}
\frac{dA}{dt}=\gamma\cdot g(A) - \delta \cdot A,
\label{eq:macroActivator}
\end{gather}
where $\gamma$ is the fully activated rate of protein synthesis and $\delta$ is the protein degradation rate (which in prokaryotes 
is often estimated from the growth rate due to growth-mediated dilution). 

In the deterministic limit, when the number of reactant molecules is very large, we expect Eq.~\ref{eq:macroActivator} to 
adequately describe the system behavior. Once initial transients have died out, the system will approach a steady-state, and $A$ 
reaches its steady-state value $A_s$ where the rate of synthesis and degradation balance, {\it i.e.} $\gamma\cdot 
g(A_s)=\delta\cdot A_s$. The stability of the steady-state is determined by the response of the system to a small perturbation 
$A_{p}$, found by linearizing Eq.~\ref{eq:macroActivator} about $A_s$,
\begin{gather}
\frac{{dA_{p} }}
{{dt}} = \left[ {\gamma  \cdot g'\left( {A_s } \right) - \delta } \right] \cdot A_{p}  \equiv \lambda  \cdot A_{p}.
\label{eq:lambdaEq}
\end{gather}
The expression in the square brackets $\lambda  \equiv \left[ {\gamma  \cdot g'\left( {A_s } \right)  - \delta } \right]$  is a 
constant that depends upon the model parameters. If $\lambda$ is positive, the small perturbations will grow in time ($A_s$ is an 
\emph{unstable state}), while if $\lambda$ is negative, the small perturbation will decay ($A_s$ is a \emph{stable} state). In the 
stable case, the long-term state of the system can be thought of as a point located at the bottom of a valley (or basin of 
attraction) -- the more negative the constant $\lambda$, the steeper the valley. As the model parameters are varied, the valley 
may become more flat ($\lambda \approx 0$) or even develop into a mountain ($\lambda > 0$), resulting in a loss of stability. The 
parameter space is divided into regions of different qualitative behavior (as in Figure 2a, black curve); the threshold between 
these domains indicates where $\lambda$ has changed sign and is called the \emph{phase boundary}. Although the model seems to 
depend upon a large family of parameters ($\gamma, \delta, K_A$, {\it etc.}), the stability of the deterministic model is actually 
described by two dimensionless combinations of these parameters: the ratio of the protein concentration with fully activated 
promoter ($A_0=\gamma/\delta$) to the dissociation constant, $A_0/K_A$, and the fold-activation, $f$.

The effective stability approximation (ESA) we propose is an approximation that allows the average effect of intrinsic noise to be 
expressed as a {\it positive} correction to $\lambda$,
\begin{gather}
\lambda' = \lambda  + \lambda _{corr} \quad \left( {\lambda _{corr}  > 0} \right),
\end{gather}
(see Eq.~\ref{eq:pertEigVals} below). The correction reflects an effective \emph{flattening} of the local landscape by stochastic 
fluctuations, making it easier for the system to escape from the basin of attraction. Adopting this perspective allows the 
analysis used to study the deterministic model to be extended to the stochastic model with only minor modification. With 
$\lambda'$ corrected to include the effect of the intrinsic noise, the new phase boundaries are drawn to coincide with points in 
parameter space where $\lambda'=0$.

A major source of intrinsic noise in gene regulatory networks is so-called translational bursting~\cite{Ouden,Kaern}, where each 
mRNA transcript is translated into several peptides before the message is degraded, leading to a \emph{burst} of protein 
synthesis. Typical values of the `burst size' $b$ can vary from close to zero for poorly translated genes~\cite{OudenBurst}, up to 
several dozen~\cite{Kennell,Xie} depending upon the rate of translation and the lifetime of the transcript. When intrinsic noise 
is included in the autoactivator model, and the procedure described in detail in \mbox{Section III--A} of the {\it Supplementary 
Material} is applied, we find the correction to $\lambda$ is $\lambda_{corr}\propto \Delta_b/\lambda^2$ where, 
\begin{gather}
\Delta_b=\frac{(b+1)}{2}\frac{1}{K_A\;V_{cell}}=\frac{(b+1)}{2}\frac{1}{N_A},
\end{gather}
is a third dimensionless quantity we call the \emph{discreteness parameter}. This parameter captures the average change in protein 
number when a synthesis or degradation event occurs, scaled relative to the protein number required to initiate activation $N_A = 
K_A \times V_{cell}$, where $K_A$ is the activator dissociation constant and $V_{cell}$ is the cell volume. Increasing the 
discreteness parameter $\Delta_b$ increases the magnitude of the discrete change in activator numbers, and therefore increases the 
relative magnitude of the perturbation to the system caused by the intrinsic noise. One would expect the circuit to switch more 
readily from stable state to stable state as the magnitude of the intrinsic noise is increased, thereby reducing the average 
stability of the circuit. On the other hand, as the number of activator molecules increases $(N_A \to \infty)$, the discreteness 
parameter vanishes and the behavior of the system is fully described by the deterministic model. Thus, the discreteness parameter 
$\Delta_b$ represents a distillation of the complicated effect of intrinsic noise on the model behavior, captured in a compact 
expression that would be difficult to extract from numerical simulation data. 

\begin{figure*}
\begin{center}
\epsfig{file=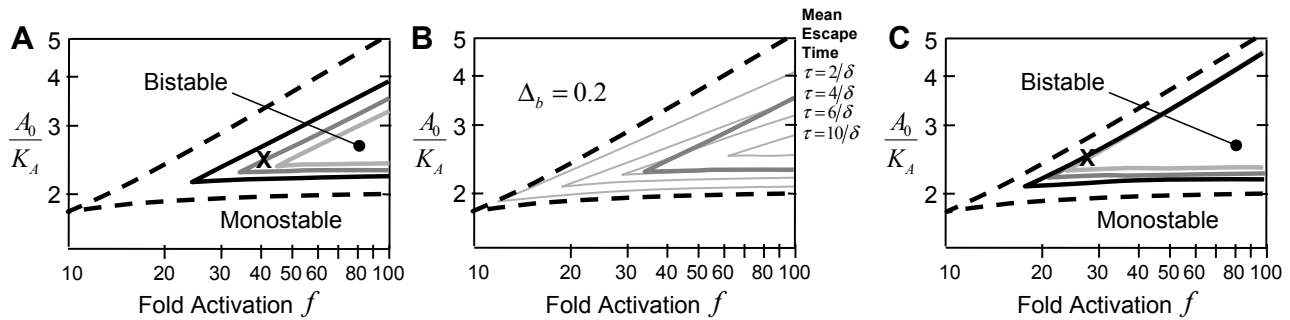,width=0.95\linewidth}
\caption{Stability phase plot for the autoactivator (Figure 1a), including the effect of intrinsic noise. (A) The black dashed 
curve is the phase boundary of the deterministic model with transcriptional activation ($A_0/K_A$ is the fully activated protein 
concentration scaled by the activator/DNA dissociation constant). Increasing the level of intrinsic noise by increasing the 
discreteness parameter $\Delta_b$ ({\it i.e.} increasing the `burstiness' of translation or decreasing the number of molecules) 
diminishes the parameter regime of reliable bistability ($Re[\lambda']<0$). Here, $\Delta_b=0.1$ (black solid), 0.2 (dark gray) 
and 0.3 (light gray). (B) The average escape time from the stable state is an indicator of the permanence of the bistability. 
Here, the dark gray curve from Figure 2a corresponds to an escape time of about $\tau=6$, where time has been scaled relative to 
the protein lifetime $\delta^{-1}$. (C) As in Figure 2a, but now with \emph{translational} activation. The range of bistability is 
considerably widened as transitions from the {\it LOW} to the {\it HIGH} state are supressed. Here, $K_A \cdot V_{cell} = 25$ 
molecules and the fully activated burst size is $b=4$ (black), $b=9$ (dark gray) and $b=14$ (light gray).}
\label{fig:AutoactivatorPhase} 
\end{center}
\end{figure*} 

As shown in Figure 2a, for the autoactivator the parameter space is divided into regions of bistability (two stable states) and 
monostability (one stable state). The bistability is most easily lost near the phase boundary separating the bistable and 
monostable states. The circuit parameters of Isaacs and co-workers~\cite{Autoactivator} lie close to the left-hand tip of the 
black triangle in Figure 2a ($f \approx 10$), and as they observed in their experiments, the noise overwhelms bistability in such 
a system  ({\it c.f.} Figure 2A of~\cite{Autoactivator}), leading to rapid transitions between the stable states. A much greater 
fold-activation is required to maintain two distinct stable states (as likewise noted by the authors). 

Actually, once noise is allowed in the autoactivator model, one no longer has stability in the strictest sense because there is 
always a chance that a perturbation will switch the system from one steady-state to the other. With noise, it is not a question of 
stability, but rather the average escape time from the steady-state~\cite{Sneppen,Wolynes}. The longer the escape time (compared 
with other time scales in the problem), the more `stable' the system. To emphasize the effect of the intrinsic noise on the 
stability phase plot, we consider a system with a small number of activator proteins $(K_A\cdot V_{cell} = 25$ molecules). Using 
the parameters $\gamma=2 \;\mbox{protein} \;\mbox{min}^{-1}$, $\delta^{-1}=30 \;\mbox{min}$ (a half-life of $\sim 
20\;\mbox{min}$), $K_A=25\; nM$ and a burst size of $b=10$, the discreteness parameter in {\it E. coli} ($V_{cell} \approx 1 \mu 
m^3$) is $\Delta_b \approx 0.2$. From Figure 2a (dark gray curve), a maximum fold-activation of $f \ge 40$ is necessary to ensure 
long-lived bistable states (shown as a cross on the plot). It is possible to explicitly compute the average escape time from the 
stable states for this simple model (see \cite{GardinerBOOK,Kepler} and \mbox{Section II} of the {\it Supplementary Material}). 
Figure 2b compares the average escape time as a function of $A_0/K_A$ and $f$ for the case above, with $\Delta_b=0.2$ (dark gray 
curve in Figure 2a). Along the dark gray curve, the escape time is $\tau=(3 \pm 0.5) \;h$, which is about six times longer than 
the protein lifetime (which sets the basic time scale of the system's `memory'). 

The escape time is an indirect measure of the system's stability. We have developed a more direct method that measures the 
effective rate of divergence of an ensemble of stochastic trajectories. This method is of general applicability and allows a 
direct evaluation of the accuracy of the ESA. The details of that calculation are reserved for the {\it Supplementary Material} 
(see Section III-A.2). Comparing $\lambda'$ to the effective rate of divergence in the stochastic simulations of the 
autoactivator, the ESA is found to be accurate for systems with $\Delta_b \lesssim 0.25$.

The burst size $b$ can be reduced by decreasing the rate of translation and indeed Ozbudak {\it et al.} suggest that many poorly 
translated genes in {\it E. coli} could be the result of evolutionary selection against burst noise~\cite{OudenBurst}. 
Alternatively, the method of control in the circuit can be shifted from transcriptional to \emph{translational} activation. 
Although the simple deterministic model remains unchanged for either choice of trancriptional or translational control, the 
resulting stochastic model exhibits improved stability for translational activation.

Figure 2c shows the result of putting the translation rate under control of the activator. Decreasing the translation rate in the 
{\it LOW} state has the effect of shifting the upper branch of the phase boundary, indicating a decrease in transitions from the 
{\it LOW} to the {\it HIGH} state. The translational autoactivator can tolerate a larger range of transcription rates ({\it i.e.} 
higher $\gamma$) and a lower maximum fold-activation ($f \ge 20$), even for large burst size. As above, with $\gamma=2 
\;\mbox{protein} \;\mbox{min}^{-1}$, $\delta^{-1}=30 \;\mbox{min}$, $K_A=25 \;nM$, $V_{cell}=1 \;\mu m^3$ and a fully activated 
burst size $b=10$, a fold-activation of $f \ge 25$ is required to sustain the bistability (shown as a cross on the plot), almost 
half that required in the transcriptional autoactivator above.

To generate the phase plot for a given stochastic model requires division of the parameter space of the model 
(Eq.~\ref{eq:macroActivator}) into a fine grid, with stochastic simulation performed at each point. Even after several such 
simulations are generated, it is unlikely that the discreteness parameter $\Delta_b$ will suggest itself as a key measure of the 
magnitude of intrinsic noise. The ESA method provides not only a rapid overview of the parameter space, but provides compact 
expressions characterizing the effect of intrinsic noise on the observable dynamics. In the next section, we shall apply ESA to 
the analysis of a more elaborate circuit model.

\subsection{Genetic oscillator}

Oscillating systems underlie many physiological processes in the cell, from circadian rhythms~\cite{GoldbeterBOOK} to the cell 
cycle itself~\cite{CellCyclePomerening}. In addition to the natural systems, several synthetic genetic oscillator designs have 
been studied, including the mutually-repressing ring-oscillator (\emph{Repressilator}) of Elowitz and Leibler~\cite{Elowitz} and 
the activator-repressor design of Atkinson and co-workers~\cite{Atkinson} (which has a great deal in common with the model 
discussed below). A recurring motif in experimentally characterized networks is a negative feedback loop serving as a system 
reset~\cite{GoldbeterBOOK,DunlapReview}. Without some time delay or intervening mechanism to prevent reversibility, the system 
will rapidly approach an intermediate equilibrium, and it is found both theoretically~\cite{CellCycleSontag} and 
experimentally~\cite{CellCyclePomerening} that a negative feedback loop alone is not sufficient to maintain reliable oscillations. 
If, however, the feedback repressor is controlled by a bistable autoactivator, the oscillations become more robust and coherent 
since the bistable switch acts as a ratchet that `locks' into the {\it HIGH} state generating a large amount of repressor to feed 
back and reset the system to the {\it LOW} state where the system remains until the activator accumulates over a critical 
threshold to initiate another cycle~\cite{CellCycleSIGGIA}. This motif is highly represented in natural gene 
networks~\cite{DunlapReview}, and we shall use the ESA to ascertain the contribution of intrinsic noise to the performance of such 
an oscillator.

We consider the generic model proposed by Vilar and co-workers to describe circadian rhythms in eukaryotes~\cite{Vilar}, with a 
transcriptional autoactivator driving expression of a repressor that provides negative control by sequestering activator proteins 
through dimerization~\cite{Constructive,TwoComponent}. The repressor and activator form an inert complex until the activator 
degrades, recycling repressor back into the system. In their model, the degradation rate of the activator, $\delta_A$, is the same 
irrespective of whether it is bound in the inert complex or free in solution. We simplify their original model somewhat, and as in 
the previous section, we assume fast activator/DNA binding and rapid mRNA turnover, leading to a reduced set of rate equations 
governing the concentration of activator $A$, repressor $R$ and the inert dimer $C$,
\begin{eqnarray}
\frac{{dA}}{{dt}} & = & \gamma _A  \cdot g\left( {\frac{A}{{K_A }},f_A } \right)  - \delta _A  \cdot A - \kappa _C  \cdot A \cdot 
R \notag\\ 
 \frac{{dR}}{{dt}} & = & \gamma _R  \cdot g\left( {\frac{A}{{K_R }},f_R } \right)  - \delta _R  \cdot R - \kappa _C  \cdot A \cdot 
R + \delta _A  \cdot C \notag\\ 
 \frac{{dC}}{{dt}} & = & \kappa _C  \cdot A \cdot R - \delta _A  \cdot C .
 \label{eq:redVilar}
\end{eqnarray}
We further assume no cooperativity in activator binding ($n=1$ in the activation function $g$) and the nominal parameter set used 
in~\cite{Vilar}. For this more complicated system, there is a larger number of dimensionless combinations of parameters that 
characterize the system dynamics. The scaled repressor degradation rate $\epsilon=\delta_R/\delta_A$ is a key control parameter in 
the model since oscillations occur in the deterministic system only for an intermediate range of this parameter. For the nominal 
parameter set used in ~\cite{Vilar}, the deterministic model exhibits oscillations over the range $0.12 < \epsilon < 40$ (Figure 
3a, black region). We shall focus on the parameter regime near to the phase boundary at $\epsilon \approx 0.12$ and examine the 
role intrinsic noise plays in generating regular oscillations from a deterministically stable system.

Applying the ESA to the oscillator model, the parameter  
$\Delta_{b_A}=(b_A+1)/(2\cdot K_A\cdot V_{cell})$ emerges as an important measure quantifying the discreteness in activator 
synthesis (see Eq. 36 in the {\it Supplementary Material}). Here again, $b_A$ is the burst size in the activator synthesis, $K_A$ 
is the activator/DNA dissociation constant and $V_{cell}$ is the cell volume. (Here, $V_{cell}=100 \mu m^3$ as is appropriate for 
eukaryotic cells.) 

\begin{figure*}
\begin{center}
\epsfig{file=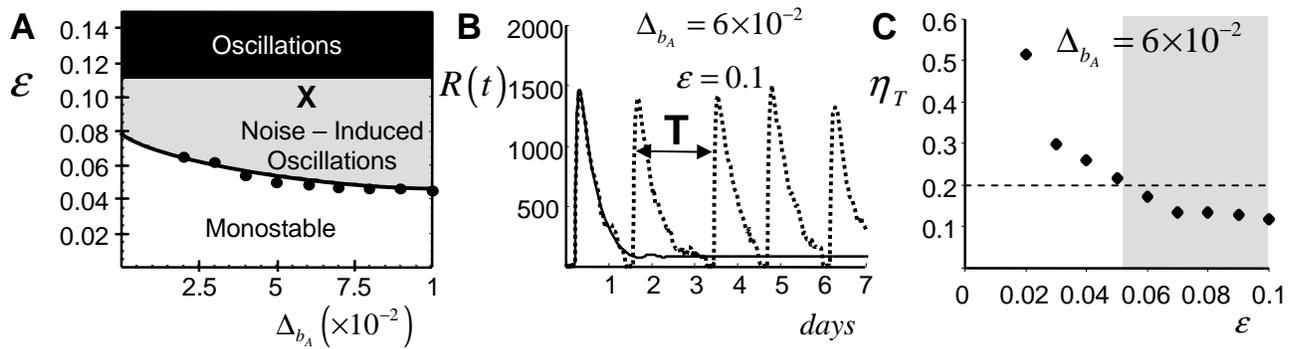,width=0.95\linewidth}
\caption{(A) Stability phase plot as a function of the scaled repressor degradation rate  $\epsilon=\delta_R/\delta_A$ for the 
circuit shown in Figure 1b. The discreteness in the activator synthesis, $\Delta_{b_A}$, characterizes the average discrete change 
in activator concentration during each reaction, and consequently the magnitude of the intrinsic noise. The intrinsic noise 
expands the region of instability (gray) extending the parameter range over which oscillations are expected to occur. The 
deterministic phase boundary is located at $\epsilon \approx 0.12$ (dashed line separating the black and gray regions). The solid 
line is the phase boundary predicted from the roots of Eq.~\ref{eq:resolve} and filled circles denote the phase boundary found by 
stochastic simulation (see text). The model and parameters are as in Vilar {\it et al.}~\cite{Vilar}. (B) The circuit exhibits 
noise-induced oscillations (dotted line) with inter-spike time $T$. The parameters used in the simulation correspond to a 
\emph{deterministically stable} system (black line). Numerical simulation data was generated using Gillespie's direct 
method~\cite{Gillespie}, with parameters as used in~\cite{Vilar} and $\epsilon =0.1, \Delta_{b_A}=6\times 10^{-2}$ (cross in 
Figure 3b). (See \mbox{Section III--C} of the {\it Supplementary Material}.) (C) A plot of the noise-to-signal ratio $\eta 
_T=\langle{ \left({\langle T \rangle-T}\right)^2}\rangle^{1/2}/\langle T \rangle$ as a function of $\epsilon$. The oscillations 
are regular when $\eta_T$ is small (the region of noise-induced oscillations predicted by the ESA is gray), and $\eta_T$ was 
calculated using at least 200 spikes for each point.}
\label{fig:vilar}
\end{center}
\end{figure*}

Using the nominal parameter set of Vilar {\it et al.}~\cite{Vilar} in our reduced model leads to a burtiness in activator 
synthesis of $b_A=5$ (giving $\Delta_{b_A}=6 \times 10^{-2}$) and a burstiness in repressor synthesis of $b_R=10$. The phase 
boundary predicted by the ESA is shown as a solid line in Figure 3a, bounding a region of parameter space between the 
deterministic phase boundary where qualitatively different behavior is expected from the stochastic model. We examine the system 
behavior in this region by running a stochastic simulation using the parameter choice $\epsilon=0.1$ and $\Delta_{b_A}=6\times 
10^{-2}$ (denoted by a cross in Figure 3a). With this choice, the deterministic model is stable (Figure 3b, black line). 
Nevertheless, a stochastic simulation of the same model, including protein bursting and stochastic dimerization, clearly shows 
oscillations (Figure 3b, dotted line). 

The time between successive peaks in the stochastic simulation of Figure 3b is denoted by $T$. As is clear from Figure 3b, $T$ is 
itself a random variable. Each simulation run generates a collection of inter-spike times from which the mean $\langle T \rangle$ 
and the variance $\langle{ \left({\langle T \rangle-T}\right)^2}\rangle^{1/2}$ can be calculated. Following Steuer {\it et 
al.}~\cite{Constructive}, the \emph{quality} of the noise-induced oscillations is measured using the noise-to-signal ratio 
$\eta_T=\langle{ \left({\langle T \rangle-T}\right)^2}\rangle^{1/2}/\langle T  \rangle$, and the system is said to exhibit regular 
oscillations where $\eta_T$ is small~\cite{Constructive,TwoComponent}. The dependence of $\eta_T$ on the repressor degradation 
rate $\epsilon$ is shown in Figure 3c, with the discreteness parameter $\Delta_{b_A}=6 \times 10^{-2}$ (as in Figure 3b), using at 
least 200 spikes to calculate $\eta_T$. At low repressor degradation rate, the noise-to-signal ratio is high, indicating large 
variance in the inter-spike time $T$ and corresponding to a \emph{stable} ({\it i.e.}, non-oscillatory) system. As the repressor 
degradation rate is increased, the variance in the inter-spike time $T$ decreases with a consequent decrease in the 
noise-to-signal ratio $\eta_T$, indicative of a more regularly oscillating system. Physically, the intrinsic noise in this 
parameter range is sufficient to drive the system away from the deterministically stable steady-state, yet the noise is not so 
strong that the return trajectory through phase space is much affected.

As in the autoactivator model, it is useful to compare the phase boundary predicted by the ESA to some independent measure of 
stability, in this case $\eta_T$. In Figure 3c, the ESA phase boundary (for $\Delta_{b_A}=6 \times 10^{-2}$)  is denoted by the 
interface between the white and gray regions, corresponding to a value of $\eta_T \approx 0.2$. Using data such as that shown in 
Figure 3c, the points in the phase plot with $\eta_T=0.2$ can be found for a range of discreteness parameter $\Delta_{b_A}$ 
(Figure 3a, filled circles). These points correspond very well to the phase boundary calculated using the ESA (Figure 3a, solid 
line). The results are as one would expect -- near the deterministic phase boundary, very little molecular noise is required to 
sustain oscillations, and reasonable periodicity persists even for small values of the discreteness parameter ($\Delta_{b_A}\to 
0$, $b_R\neq 0$). As the repressor degradation rate $\epsilon$ is decreased to a region favoring stability, more noise is required 
to overcome the deterministic stability of the system and initiate the autoactivator trigger. It is illustrative to remark that 
each data point in Figure 3a, obtained from stochastic simulation~\cite{Gillespie}, took roughly a day to generate on a dual 
processor desktop computer since at low repressor degradation rate, a large separation of timescales is introduced necessitating 
long stochastic simulation runs to capture the slowly-varying dynamics of the system. By contrast, the solid line generated from 
the roots of Eq.~\ref{eq:resolve}, took less than an hour to produce on the same machine.  Thus, even for a two-gene circuit with 
several degrees of freedom, the ESA affords a compact and convenient means to survey the phase space, drawing attention to those 
regions of particular interest that may be probed in more detail by more realistic (though also more computationally costly) 
stochastic simulation methods. 

\section{Methods}

The effective stability approximation can be applied to generic models of genetic circuits in a straightforward way. Here, a brief 
outline of the method is provided. A self-contained tutorial on stochastic modeling and the ESA is found in the {\it Supplementary 
Material}.

A useful abstraction of genetic regulatory networks is as a system of ordinary differential 
equations~\cite{ConradTysonBOOK,KaernsWeissBOOK}. (Here, and throughout, we shall assume a spatially homogeneous environment.) We 
denote the \emph{concentrations} of the reactants of interest by the state vector ${\bf x}$, where the $x_i$ correspond to the 
concentration of mRNA, transcription factors, protein products, {\it etc}. The kinetic equation governing the evolution of the 
system takes the form $\frac{d{\bf x}}{dt}={\bf f}({\bf x})$, where ${\bf f}$ is a vector of nonlinear functions of the state 
variables. We can estimate the long-time, or steady-state, behavior of the model by first computing the equilibrium points ${\bf 
x}_s$ that satisfy the algebraic constraint ${\bf  f}({\bf x}_s)={\bf 0}$. We then Taylor expand the reaction rate vector about 
the equilibrium point by making the substitution ${\bf x}={\bf x}_s+{\bf x}_{p}$ (where ${\bf x}_{p}$ is an infinitesimal 
perturbation away from ${\bf x}_s$), and retain only linear terms in ${\bf x}_{p}$. The resulting dynamics of ${\bf x}_{p}$ are 
given by$\frac{d}{dt}{\bf x}_{p}={\bf J}\cdot {\bf x}_{p}$, where ${\bf J}$ is the Jacobian or response matrix: ${\bf 
J}_{ij}=\partial f_i/\partial x_j$. The eigenvalues of ${\bf J}$ are the matrix analogue of the parameter $\lambda$ introduced in 
Eq.~\ref{eq:lambdaEq}, and in a similar fashion if the eigenvalues all have negative real-part, then ${\bf x}_s$ is a \emph{stable 
steady state}. (There are, of course, limitations to how far one can trust the linearization~\cite{StrogatzBOOK}, but for our 
purposes it is sufficient as a first approximation.)

To include stochastic effects in the mathematical model, chemical reaction rates must be re-written in terms of the reaction 
\emph{propensity} and \emph{stoichiometry}~\cite{Gillespie}. For example in the positive autoactivator example above, with the 
individual synthesis and degradation stoichiometries written explicitly, the deterministic model equations 
(Eq.~\ref{eq:macroActivator}) read,
\begin{gather}
\begin{array}{*{20}c}
   {\mbox{bursty synthesis:}} & {A\xrightarrow{{\nu _1 }}A + b;} & {\nu _1  = \frac{{\gamma}}
{b}\cdot g \left( A \right)},  \\
   {\mbox{linear degradation:}} & {A\xrightarrow{{\nu _2 }}A - 1;} & {\nu _2  = \delta\cdot A.}  \\
 \end{array}
 \label{eq:tslRates} 
\end{gather}
We encode this information concisely as the \emph{propensity vector} $\bm {\nu} = \lbrack \nu_1, \nu_2 \rbrack = \lbrack{ \gamma 
\cdot g(A)/b, \delta\cdot A }\rbrack$ and the stoichiometry matrix ${\bf S} = \lbrack b, -1 \rbrack$. The discrete change in 
molecule numbers following the completion of a chemical reaction causes a deviation from the deterministic solution since the 
deterministic model assumes an infinitesimally small and \emph{continuous} change in the state. (Consequently, the deterministic 
model only applies to systems with large numbers of molecules.) We denote the deviation of the stochastic model from the 
deterministic model by the fluctuating quantity $\omega\cdot \bm{\alpha}(t)$, where $\omega = 1/ \sqrt{V_{cell}}$ and 
$\bm{\alpha}(t)$ describes the stochastic deviation in each species ${\bf x}$. The $\sqrt{V_{cell}}$ scaling arises from the 
observation that the relative magnitude of the intrinsic noise scales roughly as the inverse square-root of the number of 
molecules~\cite{vanKampenME}. Elf and Ehrenberg~\cite{Elf} have developed an algorithmic expression for the statistics of 
$\bm{\alpha}$ using the linear noise approximation of van Kampen~\cite{vanKampenME}. In that formulation, the mean and covariance 
of the fluctuations about the deterministic state are written compactly in terms of the propensity vector $\bm{\nu}$ and the 
stoichiometry matrix ${\bf S}$; here, we shall apply their method to characterize the fluctuations about the stable state. The 
first step in the calculation of the moments of the fluctuations $\omega \bm{\alpha}(t)$ is to construct the auxiliary matrices 
${\bf \Gamma}$ and ${\bf D}$, evaluated at the stable state ${\bf x}_s$,
\begin{gather}
\Gamma_{ij}(t) = \frac{\partial \lbrack{{\bf S}\cdot {\bm \nu}}\rbrack_i}{\partial x_j}=\frac{{\partial  
f_i }}{{\partial x_j }} \quad \quad{\bf D}={\bf S}\cdot \mbox{diag}[{\bm \nu}]\cdot {\bf S}^T.
\label{eq:elfMat}
\end{gather}
The drift matrix ${\bf \Gamma}={\bf J}$ is the response matrix (or Jacobian) described above and reflects the local stability of 
the deterministic system to small perturbations~\cite{LimitCycle1}. The diffusion matrix ${\bf D}$ captures the strength of the 
fluctuations and is related to the magnitude of the reaction step-size~\cite{Elf,Chaos}. It is straightforward to show that to 
leading-order in $\omega$ the mean of the fluctuations is zero ($\langle \bm{\alpha} \rangle=\bm{0}$) and the variance, denoted by 
the symmetric matrix ${\bm \Xi}=\langle{\bm \alpha}\cdot{\bm \alpha}^T\rangle$, is determined by the solution of the system of 
algebraic equations~\cite{vanKampenME},
${\bf{\Gamma}} \cdot {\bf{\Xi}}  + {\bf{\Xi}}  \cdot {\bf{\Gamma}}^T  + {\bf{D}} = {\bf 0}$. Since the fluctuations about the 
stable state are stationary, the time autocorrelation function depends upon the time difference only, and is given by the matrix 
exponential,
\begin{gather}
\left\langle {{\bm{\alpha }}\left( t \right){\bm{\alpha }}^T \left( {t - \tau } \right)}  
\right\rangle  = \exp \left[ {{\bf{\Gamma}}\tau } \right] \cdot {\bf{\Xi}}.
\label{eq:timeCorrfun}
\end{gather}
The effect of the fluctuations on the deterministic steady-state is calculated by including an additional term in the 
deterministic linearization above: ${\bf x}={\bf x}_s+{\bf  
x}_{p}+\omega{\bm \alpha}$. Linearizing ${\bf J}$ in $\omega$, we have a stochastic differential equation governing the decay of 
the perturbation modes ${\bf  x}_{p}$,
\begin{gather}
\frac{d}{dt}{\bf x}_{p}=\lbrack{{\bf J}^{(0)}+\omega\;{\bf J}^{(1)}(t)}\rbrack \cdot {\bf x}_{p}.
\end{gather}
The fluctuations affect the decay of the infinitesimal disturbance ${\bf x}_{p}$ as well as the dynamics of the average $\langle 
{\bf x}_{p} \rangle$, which (provided $\omega\;{\bf J}^{(1)}(t) \ll {\bf J}^{(0)}$) is approximately governed by the convolution 
equation~\cite{Bourret2,vanKampenStochDE},
\begin{gather}
\frac{d}{{dt}}\left\langle {{\bf x}_{p}\left( t \right)} \right\rangle  = {\bf  
J}^{(0)}\left\langle {{\bf x}_{p}\left( t \right)} \right\rangle  + 
\omega ^2 \int\limits_0^t {{\mathbf{J}}_c \left( t-\tau  \right)\left\langle {{\bf x}_{p}\left(  
{\tau } \right)} \right\rangle d\tau },
\end{gather}
where ${\mathbf{J}}_c \left( t-\tau  \right)=\left\langle {{\mathbf{J}}^{\left( 1 \right)} \left( t  
\right)e^{{\mathbf{J}}^{\left( 0 \right)} (t-\tau)} {\mathbf{J}}^{\left( 1 \right)} \left( \tau  
\right)} \right\rangle$ is made up of linear combinations of the cross-correlations $\langle  
\alpha_i(t)\alpha_j(\tau)\rangle$ given by the $i^{th}$ row and the $j^{th}$ column of the right-hand side of 
Eq.~\ref{eq:timeCorrfun}. In the noiseless case, the stability of the perturbation ${\bf x}_{p}$ is determined by the eigenvalues 
of ${\bf J}^{(0)}$:  
\mbox{$\mbox{diag}\lbrace \lambda_i \rbrace = {\bf P}^{-1}\cdot {\bf J}^{(0)}\cdot {\bf P}$} where the matrix ${\bf P}$ is made of 
the eigenvectors of ${\bf J}^{(0)}$. The analogues of the eigenvalues for the convolution equation above are found from the poles 
of the Laplace transform, denoted $\lambda'$, which solve the resolvent equation~\cite{VolterraStability2},
\begin{gather}
\det \left[ {\lambda '{\mathbf{I}} - {\mathbf{J}}^{\left( 0 \right)}  - \frac{1}{V_{cell}}  
{\mathbf{\hat J}}_c \left( {\lambda '} \right)} \right] = 0,
\label{eq:resolve}
\end{gather}
Here $\omega^2$ has been replaced by $V^{-1}_{cell}$ and ${\mathbf{\hat J}}_c \left( s \right) =  
\int\limits_0^\infty  {{\bf J}_c(t) e^{ - st} dt} $ is the Laplace transform of ${\bf J}_c(t)$. If the deterministic eigenvalues 
are distinct, we can further approximate the effective eigenvalue $\lambda'_{i}$ by,
\begin{gather}
\lambda'_{i} = \lambda_i + \frac{1}{V_{cell}} \;[\;{\bf P}^{-1} \cdot {\mathbf{\hat J}}_c \left(  
\lambda_i \right) \cdot {\bf P}\;]_{ii}.
\label{eq:pertEigVals}
\end{gather}
where $[\;\cdot\;]_{ii}$ denotes the $i^{th}$ diagonal entry of the matrix. Physically, we interpret the leading-order noise 
correction as the \emph{power} in the fluctuations at eigenfrequency $\lambda_i$ projected in the eigendirection of $\lambda_i$. 
Since the correction term is quadratic, it is always positive and thus {\it de}-stabilizes the eigenmode upon which it is 
projected. (Hence, in Eq. 4 we write $\lambda_{corr} > 0$.)

It often happens that out of the term $1/V_{cell} \;[\;{\bf P}^{-1} \cdot {\mathbf{\hat J}}_c \left(  \lambda_i \right) \cdot {\bf 
P}\;]_{ii}$ there appears a small parameter that quantifies the effect of the intrinsic noise. (For the two examples above, the 
small parameters are $\Delta_b$ and $\Delta_{b_A}$, each characterizing the \emph{discreteness} of the protein change.) In the 
limit that this parameter goes to zero, the effect of the intrinsic noise becomes negligible, at least in that particular 
eigenmode. 

Finally, the ESA can be easily implemented in a symbolic computational environment, without attending to the mathematical details 
(see Section IV of the {\it Supplementary Material}). A version of the ESA coded in {\it Mathematica} is freely available from the 
authors by request.

\begin{acknowledgments}
The authors thank Jian Liu, Francis Poulin and Stefan Klumpp for critical reading and constructive comments on the manuscript. MS 
is grateful for the post-doctoral fellowship funding provided by Canada's NSERC. BI is supported by an NSERC Discovery grant. This 
work was supported in part by NSF Grant No. MCB0417721 through TH, and by Grant No. PHY-0216576 and PHY-0225630 through the 
PFC-sponsored Center for Theoretical Biological Physics.
\end{acknowledgments}

\newpage

\section{Supplementary Material}

Much theoretical work has been devoted to quantifying the conditions under which microscopic fluctuations have macroscopic 
effects~\cite{NoiseBOOK}. The most useful results are often restricted to systems with a single degree of freedom or employ 
sophisticated tools such as It\^o's calculus. In what follows, we aim to develop a convenient and simple scheme to assess the 
stability properties of a dynamical system subject to molecular noise described by the chemical Master equation. The method is an 
extension of the familiar linear stability analysis of nonlinear dynamical systems, although here the effective eigenvalues about 
the equilibrium points are adjusted to reflect the influence of the noise.

\section{Mathematical Methods}

A very useful qualitative picture of the behavior of a system of nonlinear differential equations emerges from the linearized 
dynamics about the \emph{fixed-point(s)} (also called the \emph{steady-state(s)}) of the system, defined as the reactant 
concentrations at which the synthesis and degradation rates balance.  The stability of the system near the fixed-points can be 
estimated by calculating the \emph{eigenvalues} $\lbrace \lambda_i \rbrace$ of
the resulting linearization, which are generally a set of complex numbers. If the real parts are all negative, we say the system 
is locally stable, meaning small perturbations away from the steady-state are automatically corrected. 

Since genetic circuits, both natural and engineered, rely upon transfer of information through small numbers of molecules, 
significant fluctuation is simply one of the inherent operating conditions~\cite{Kerszberg}, resulting in noise that may give rise 
to behavior that is very different from the behavior predicted by deterministic models. Consequently, for cell-scale modeling we 
propose to modify the deterministic notion of stability by calculating the \emph{effective} eigenvalues $\lambda'_i$, which 
include the averaged influence of the intrinsic noise,
\begin{gather}
\lambda_i'=\lambda_i + \lambda_{corr}.
\end{gather}
Here $\lambda_{corr}\propto V_{cell}^{-1}$ is inversely proportional to the cell volume $V_{cell}$ For notational convenience in 
the following, we introduce a parameter $\omega$ that is related to the cell volume by: $\omega^{-2}=V_{cell}$. Sometimes 
$\omega^{-2}$ is called the `system size', expressing as it does the relationship between reactant concentration and molecule 
numbers~\cite{Elf,vanKampenBOOK}. 

\subsection{Stochastic stability equation}

To calculate the stability of the macroscopic model $\frac{d{\bf x}}{dt}={\bf f}({\bf x})$ to small perturbations, the system is 
linearized about the equilibrium point: ${\bf x}={\bf x}_{s}+{\bf x}_{p}$, 
\begin{gather}
\frac{d}{{dt}}{\bf{x}}_{p} = {\bf{J}^{(0)}} \cdot {\bf{x}}_{p}.
\label{eq:noiselessmode}
\end{gather}
(Here, and henceforth, we adopt the convention of writing all matrix variables in bold upper-case, and all vectors in bold 
lower-case.) The eigenvalues of the Jacobian ${\bf{J}}^{(0)} = \left. {\frac{{\partial {\bf{f}}}}{{\partial {\bf{x}}}}} 
\right|_{{\bf{x}} = {\bf{x}}_{s} } $ provide the decay rate of the exponential eigenmodes; if all the eigenvalues have negative 
real part, we say the system is \emph{locally asymptotically stable}. We shall restrict ourselves to this notion of stability, 
although it does ignore \emph{algebraically} growing modes which may be important in certain instances \cite{TrefethenBOOK}.

To accommodate fluctuations on top of the small perturbation ${\bf x}_{p}$, we set ${\bf x} = {\bf x}_{s}+{\bf x}_{p} + \omega 
{\bm \alpha}(t)$. The Jacobian
\begin{gather*}
{\bf J} \equiv  \left. {\frac{{\partial {\bf{f}}}}{{\partial {\bf{x}}}}} \right|_{{\bf{x}} = {\bf{x}}_{s}  + \omega{\bm{\alpha }} 
},
\end{gather*}
will then be a (generally) nonlinear function of the fluctuations about the steady-state ${\bm \alpha}(t)$. (As a technical aside, 
we note that we are justified in replacing ${\bf x}$ by ${\bf x}_s+{\bf x}_{p}+\omega \bm{\alpha}(t)$ in both the right- 
\emph{and} left-hand side of the deterministic model $\frac{d{\bf x}}{dt}={\bf f}({\bf x})$ since the fluctuations 
$\bm{\alpha}(t)$ have non-zero correlation time (as we show below) and zero mean, allowing us first to conclude that the 
time-derivative of $\bm{\alpha}(t)$ exists and further that the average of this derivative must vanish: 
$\langle{\frac{d\bm{\alpha}}{dt}}\rangle=\frac{d\langle\bm{\alpha}\rangle}{dt}=0$). In the limit $\omega \to 0$, we can further 
linearize ${\bf J}$ with respect to $\omega$,
\begin{gather*}
{\mathbf{J}} \approx \left. {\mathbf{J}} \right|_{\omega  \to 0}  + \omega \left. {\frac{{\partial {\mathbf{J}}}}
{{\partial \omega }}} \right|_{\omega  \to 0}  \equiv {\mathbf{J}}^{(0)}  + \omega {\mathbf{J}}^{(1)} \left( t \right).
\end{gather*}
The stability equation is then given by,
\begin{gather}
\frac{d}{{dt}}{\bf{x}}_{p} =[ {\bf{J}^{(0)}} + \omega {\bf J}^{(1)} (t) ]\cdot {\bf{x}}_{p}.
\label{eq:stochstab}
\end{gather}
This is a linear stochastic differential equation with random coefficient matrix ${\bf J}^{(1)} (t)$ composed of a linear 
combination of the steady-state fluctuations ${\bm \alpha} (t)$ which have \emph{non-zero} correlation time ({\it cf.} 
Eq.~\ref{eq:timeCorrfun}). We therefore need not appeal to any specialized calculi ({\it e.g.} It\^o's calculus) for 
interpretation since the non-vanishing correlation time of the fluctuations ensures that ${\bf{x}}_{p}$ is a differentiable 
process and the equation falls under the purview of ordinary calculus~\cite{ItoVSStartonovich}.

Our present interest is in the \emph{mean stability} of the equilibrium point. Taking the ensemble average of 
Eq.~\ref{eq:stochstab},
\begin{gather*}
\frac{d}{{dt}}\left\langle {{\bf{x}}_{p}} \right\rangle  = {\bf{J}}^{(0)}  \cdot \left\langle {{\bf{x}}_{p}} \right\rangle  + 
\omega \left\langle {{\bf{J}}^{(1)} \left( t \right) \cdot  {\bf{x}}_{p}} \right\rangle .
\end{gather*}
The right-most term is the cross-correlation between the process ${\bf x}_{p}$ and the coefficient matrix ${\bf J}^{(1)} (t)$. 
Since the correlation time of ${\bf J}^{(1)} (t)$ is not small compared with the other time scales in the problem, it cannot be 
replaced by white noise, and an approximation scheme must be developed to find a closed evolution equation for $\langle {\bf 
x}_{p} \rangle$.

\subsection{Bourret's mode-coupling approximation}

By assumption, the number of molecules is large so the parameter $\omega$ is small, although not so small that intrinsic 
fluctuations can be ignored. To leading-order in $\omega$, the trajectory ${\bf x}_{p}\left(t\right)$ is a random function of time 
since it is described by a differential equation with random coefficients. Derivation of the entire probability distribution of 
${\bf x}_{p}\left(t\right)$ is usually impossible, and we must resort to methods of approximation. We shall adopt the closure 
scheme of Bourret \cite{Bourret1,Bourret2,vanKampenStochDE} to arrive at a deterministic equation for the evolution of the 
averaged process $\left\langle {{\bf x}_{p}\left( {t} \right)} \right\rangle $ in terms of only the first and second moments of 
the fluctuations. In that approximation, provided ${\bf J}^{(0)} \gg \omega {\bf J}^{(1)}$, the dynamics of $\langle {\bf 
x}_{p}\rangle$ are governed by the convolution equation,
\begin{gather}
\frac{d}{{dt}}\left\langle {{\bf x}_{p}\left( t \right)} \right\rangle  = {\bf J}_0\left\langle {{\bf x}_{p}\left( t \right)} 
\right\rangle  \\\notag
+ 
\omega ^2 \int\limits_0^t {{\mathbf{J}}_c \left( t-\tau  \right)\left\langle {{\bf x}_{p}\left( {\tau } \right)} \right\rangle 
d\tau },
\end{gather}
where ${\mathbf{J}}_c \left( t-\tau  \right)=\left\langle {{\mathbf{J}}^{\left( 1 \right)} \left( t \right)e^{{\mathbf{J}}^{\left( 
0 \right)} (t-\tau)} {\mathbf{J}}^{\left( 1 \right)} \left( \tau \right)} \right\rangle$ is the time autocorrelation matrix of the 
fluctuations and $e^{{\bf J}_0 \tau}$ is the \emph{matrix exponential}. The equation can be solved formally by Laplace transform, 
\begin{gather*}
\left\langle {{\bf{\hat x}_{p}}\left( s \right)} \right\rangle  = \left[ {s{\mathbf{I}} - {\mathbf{J}}^{(0)}  - \omega ^2 
{\mathbf{\hat J}}_c \left( s \right)} \right]^{ - 1} \left\langle {{\bf x}_{p}\left( 0 \right)} \right\rangle ,
\end{gather*}
where now ${\mathbf{\hat J}}_c \left( s \right) = \int\limits_0^t {{\mathbf{J}}_c \left( t \right)e^{ - st} dt} $. A necessary and 
sufficient condition for asymptotic stability of the averaged perturbation modes $\left\langle {{\bf x}_{p}\left( t \right)} 
\right\rangle$ is that the roots $\lambda '$ of the resolvent,
\begin{gather}
\det \left[ {\lambda'{\mathbf{I}} - {\mathbf{J}}_0  - \omega ^2 {\mathbf{\hat J}}_c \left( \lambda' \right)} \right] = 0,
\label{eq:renormJac} 
\end{gather}
all have negative real parts $(Re(\lambda')<0)$~\cite{VolterraStability1,VolterraStability2}. Some insight into the behavior of 
the system can be gained by considering a perturbation expansion of the effective eigenvalues $\lambda'$ in terms of the small 
parameter $\omega$. We further diagonalize ${\bf J}^{(0)}$, $\mbox{diag}[\lambda_i]={\bf P}^{-1} \cdot {\bf J}^{(0)} \cdot {\bf 
P}$, and provided the eigenvalues are distinct, we can explicitly write $\lambda'_i$ in terms of the unperturbed eigenvalues 
$\lambda_i$ to $O(\omega^4)$ as,
\begin{gather}
\lambda'_i = \lambda_i + \omega^2 \;[\;{\bf P}^{-1} \cdot {\mathbf{\hat J}}_c \left( \lambda_i \right) \cdot {\bf P}\;]_{ii},
\label{eq:pertEigVals_Supp}
\end{gather}
where $[\;\cdot\;]_{ii}$ denotes the $i^{th}$ diagonal entry of the matrix.

Notice the matrix product ${\bf J}_c(t-\tau)$ contains linear combinations of the correlation of the fluctuations $\langle 
\alpha_i(t)\alpha_j(\tau)\rangle$, and as such we must derive an expression for those moments. 

\subsection{Calculating the statistics of the steady-state fluctuations}

The statistics of the fluctuations $\bm{\alpha}$ are fully determined by the solution of the chemical Master equation (defined 
below) that comes from treating each reaction event probabilistically. In that probabilistic formulation, our state at any time 
$t$ is represented by the vector of molecule numbers ${\mathbf{n}} \in \mathbb{N}^d $; with $n_i$ representing the number of 
molecules of a given species. Each reaction causes a transition from the initial state ${\bf n}$ to some new state ${\bf n}'$ 
reflecting the addition or removal of molecules by that reaction. The probability that the transition $ {\bf n} \to {\bf n}'$ 
occurs is the product of the probability of being in state ${\bf n}$ at time $t$, $P({\bf n},t)$, and the transition probability 
of moving from $ {\bf n} \to {\bf n}'$, denoted by $W_{{\bf n} \to {\bf n}'}$. We thus write the probability conservation as a 
balance of flux into and out of the state ${\bf n}$, which yields a discrete differential equation for $P({\bf n},t)$,
\begin{gather}
\frac{{\partial P\left( {{\bf{n}},t} \right)}}{{\partial t}} = \sum\limits_{{\bf{n'}}} {W_{{\bf{n'}} \to {\bf{n}}} P\left( 
{{\bf{n'}},t} \right) - W_{{\bf{n}} \to {\bf{n'}}} P\left( {{\bf{n}},t} \right)}.
\label{eq:mastereqa}
\end{gather}
The evolution equation for $P({\bf n},t)$ is called the Master equation~\cite{ChemMaster}. It is rare that the Master equation can 
be solved exactly for $P({\bf n},t)$, and approximation schemes are required. One such scheme, the linear noise 
approximation~\cite{vanKampenME}, is versatile and will be described briefly (see also \cite{Elf} and \cite{Chaos}). The 
approximation begins with the assumption that the molecule concentrations can be meaningfully separated into a component that 
evolves deterministically, which we shall denote ${\bf x}(t)$, and fluctuations ${\bm \alpha}(t)$ that account for the deviation 
of the stochastic model from the deterministic model. We introduce a scaling parameter $\omega$, where $\omega^{-2}=V_{cell}$ is 
the volume of the cell and is an extensive measure of the number of molecules. We then make the ansatz that the fluctuations scale 
as the square-root of the number of molecules: $\omega^2\;n_i=x_i + \omega\;\alpha_i$~\cite{vanKampenME,KuboME}. In that way, a 
perturbation expansion as the number of molecules gets large ($\omega \to 0$, with concentration held fixed), returns to zero'th 
order the macroscopic reaction rate equations, 
\begin{equation}
\frac{d \bf{x}}{dt} = \bm{f} ( \bf{x} ).
\label{MacroEqs}
\end{equation}
The first-order equation, that comes at $O(\omega)$, characterizes the probability distribution for the fluctuations $\Pi ({\bm 
\alpha},t)$ centered on the macroscopic trajectory ${\bf x}(t)$, and has the form of a \emph{linear} Fokker-Planck equation,
\begin{equation}
\frac{{\partial \Pi }}{{\partial t}} =  - \sum\limits_{i,j} {\Gamma_{ij} {\partial_i}(\alpha _j \Pi )} 
+ \frac{1}{2}\sum\limits_{i,j} {D_{ij} \partial_{ij} \Pi }.
\label{LNAFP}
\end{equation} 
where $\partial_i$ denotes ${\partial  \mathord{\left/
 {\vphantom {\partial  {\partial \alpha _i }}} \right.
 \kern-\nulldelimiterspace} {\partial \alpha _i }}$ and
\begin{gather}
\Gamma_{ij}(t) = \frac{{\partial f_i }}{{\partial x_j }} \quad \quad \bm{D}={\bf S}\cdot \mbox{diag}[{\bm \nu}]\cdot {\bf S}^T,
\label{eq:elfMat_Supp}
\end{gather}
(see main text). The matrices $\bm{\Gamma}$ and $\bm{D}$ are independent of $\bm{\alpha}$, which appears only linearly in the 
drift term. As a consequence, the distribution $\Pi (\bm{ \alpha},t)$ will be Gaussian for all time. In particular, at equilibrium 
the fluctuations are distributed with density,
\begin{gather*}
\Pi _s \left( {\bm{\alpha }} \right) = \left[ {\left( {2\pi } \right)^d \det {\mathbf{\Xi}}} \right]^{\frac{1}
{2}} \exp \left[ { - \frac{1}
{2}{\bm{\alpha }}^T  \cdot {\mathbf{\Xi}}^{ - 1}  \cdot {\bm{\alpha }}} \right],
\end{gather*}
and variance ${\bf \Xi}=\langle{\bm \alpha}\cdot{\bm \alpha}^T\rangle$ determined by,
\begin{gather}
{\bf{\Gamma}} \cdot {\bf{\Xi}}  + {\bf{\Xi}}  \cdot {\bf{\Gamma}}^T  + {\bf{D}} = 0.
\label{eq:flucDissipation}
\end{gather}
Furthermore, the steady-state time correlation function is,
\begin{gather}
\left\langle {{\bm{\alpha }}\left( t \right){\bm{\alpha }}^T \left( {t - \tau } \right)} \right\rangle  = \exp \left[ 
{{\bf{\Gamma}}\tau } \right] \cdot {\bf{\Xi}}.
\label{eq:timeCorrfun_Supp}
\end{gather}
Around the steady-state, the process is stationary, which means the correlation function depends upon time difference only. Also 
note that the characteristic correlation time $\tau_c = || {\bf \Gamma}||^{-1}$ is related to the Jacobian ${\bf \Gamma}$ of the 
deterministic equations, and therefore \emph{cannot} be divorced from the deterministic relaxation time. As a consequence, 
representing the fluctuations ${\bm \alpha}(t)$ as white noise $(\tau_c \to 0)$ is \emph{not} justified. 

The great advantage of the linear noise approximation is that the autocorrelation function of the steady-state fluctuations can be 
calculated directly from the macroscopic reaction rates in an algorithmic fashion~\cite{Elf}. Furthermore, since ${\bf \Gamma}$ 
and ${\bf D}$ are derived from the known propensity and stoichiometry of the reactions, the statistics of $\bm{\alpha}$ are fully 
determined and are \emph{not} tunable by some {\it ad hoc} prescription.

\section{Mean first passage time}

Bistability is a property exhibited by deterministic systems. In a stochastic context, bistability is sometimes assigned to an 
equilibrium probability distribution with two maxima, irrespective of their separation. A more practical criterion for bistability 
is that the two states are long-lived and that the mean escape time from one state to the other is longer than the natural 
timescales in the problem. For the single-variable autoactivator model, we are able to compute the escape time by an explicit 
(though approximate) expression (see ~\cite{Kepler} or p. 139 of~\cite{GardinerBOOK} for details). Under fairly unrestrictive 
assumptions~\cite{GillespieChemLangevin}, the Master equation may be approximated by the nonlinear Fokker-Planck equation,
\begin{gather*}
\frac{{\partial P\left( {a,t} \right)}}
{{\partial t}} =  - \frac{\partial }
{{\partial a}}\Gamma\left( a \right)P\left( {a,t} \right) + \frac{1}
{2}\frac{{\partial ^2 }}
{{\partial a^2 }}D\left( a \right)P\left( {a,t} \right),
\end{gather*}
where the functions $\Gamma$ and $D$ are the nonlinear analogues of the coefficient matrices ${\bf \Gamma}$ and ${\bf D}$ 
generated by the linear noise approximation shown in the previous section. For our autoactivator example, the coefficients are 
given by,
\begin{gather*}
\Gamma(a)=\gamma\cdot g(a)-\delta\cdot a\quad\quad D(a)=\gamma\cdot b\cdot g(a)+\delta\cdot a.
\end{gather*}
The nonlinear Fokker-Planck equation has no general solution for systems of dimension greater than 1, and even the stationary 
solution is often impossible to calculate exactly for such systems~\cite{RiskenBOOK}. In the reduced autoactivator model, we are 
fortunate to have a system with one independent variable, so we can write the stationary solution of the Fokker-Planck equation 
explicitly as,
\begin{gather*}
P^s(a)=\frac{\mathcal{N}}{D(a)}\;\mbox{exp}\left[ {2\int\limits_0^a {\frac{{\Gamma\left( {a'} \right)}}
{{D\left( {a'} \right)}}da'} } \right],
\end{gather*}
where $\mathcal{N}$ is the constant of normalization (see p. 124 of~\cite{GardinerBOOK}). Furthermore, we can explicitly write the 
first passage time $\tau$ from the {\it HIGH} state to the {\it LOW} state or vice-versa. 
\begin{gather*}
\tau_{HI \to LO}=2\int\limits_{a_{mid} }^{a_{HI}^* } {\frac{1}
{{\psi \left( x \right)}}\int\limits_x^\infty  {\frac{{\psi \left( y \right)}}
{{D\left( y \right)}}dy} dx} \\
\tau_{LO \to HI}=2\int\limits_{a_{LO}^{\star} }^{a_{mid} } {\frac{1}
{{\psi \left( x \right)}}\int\limits_0^x {\frac{{\psi \left( y \right)}}
{{D\left( y \right)}}dy} dx},
\end{gather*}
where $a_{mid}$ is the unstable equilibrium point separating the {\it HIGH} and {\it LOW} states $a^{\star}_{HI}$ and 
$a^{\star}_{LO}$, respectively. The function $\psi (x)$ is given by,
\begin{gather*}
\psi(x)=\mbox{exp}\left[ {2\int\limits_0^x {\frac{{\Gamma\left( {x'} \right)}}
{{D\left( {x'} \right)}}dx'} } \right],
\end{gather*}
(see p. 139 of~\cite{GardinerBOOK} for additional details). 

In the main text, we discuss $\mbox{min}\lbrack \tau_{LO \to HI},\tau_{HI \to LO}\rbrack$ along the stability curves predicted by 
the effective eigenvalues. For \mbox{$\Delta_b=0.1$, $\mbox{min}\lbrack \tau_{LO \to HI},\tau_{HI \to LO}\rbrack = $}$8 \pm 4$, 
where time has been scaled to protein lifetime ($\delta^{-1}$). For $\Delta_b=0.2$ and $\Delta_b=0.3$, $\mbox{min}\lbrack \tau_{LO 
\to HI},\tau_{HI \to LO}\rbrack = 5.6 \pm 1.4$ and $5.9 \pm 0.3$, respectively.

\section{Details of Genetic Circuit Examples}

\subsection{The autoactivator}

We describe the transcription of the activator mRNA, $m_a$ and the translation of activator protein $A$ as two differential 
equations using the activation function $g$ to describe the time-averaged state of the promoter,
\begin{gather}
\frac{dm_a}{dt}=\gamma_m\cdot g(A) -\delta_m \; m_a,
\frac{dA}{dt}=\gamma_p\;m_a -\delta_p \; A.
\end{gather}
Here $\gamma_m$ is the transcription rate, $\gamma_p$ is the translation rate, $\delta_m$ and $\delta_p$ are the rates of mRNA 
degradation and protein degradation, respectively. We make the assumption that the mRNA turnover is much faster than the timescale 
of protein degradation ({\it i.e.} $\delta_m \gg \delta_p$). In that way, we justify setting the mRNA concentration to its 
equilibrium level,
\begin{gather}
m^{\star}(A)=\frac{\gamma_m}{\delta_m}\;g(A),
\end{gather}
reducing the model to a single equation,
\begin{gather}
\frac{dA}{dt}=\frac{\gamma_m\cdot\gamma_p}{\delta_m}\cdot g(A) -\delta_p \;A,
\end{gather}
at the expense of lumping transcription and translation together. Re-writing the constants 
$\gamma=\frac{\gamma_m\cdot\gamma_p}{\delta_m}$ and $\delta_p=\delta$, we are left with the evolution equation as written in the 
main text,
\begin{gather}
\frac{dA}{dt}=\gamma\cdot g(A) - \delta\cdot A,
\end{gather}
where $\gamma$ is the fully activated rate of protein synthesis and $\delta$ is the rate of protein degradation.

\subsubsection{Transcriptional activation}

The lumping together of transcription and translation comes at the expense of obscuring translational amplification of the mRNA. 
The translational burst size is approximately equal to the averaged number of protein molecules synthesized during the lifetime of 
the mRNA, $b=\frac{\gamma_p}{\delta_m}$~\cite{Kaern,Ouden}, so we see the production term in the macroscopic equation is actually 
($b \times$ transcription rate),
\begin{gather}
\frac{dA}{dt}=b \times \gamma_m\cdot g(A) -\delta \cdot A.
\end{gather}
In the deterministic model, the distinction between reaction rate and reaction stoichiometry is immaterial, but that is no longer 
true when we calculate the intrinsic fluctuations. Writing the production and degradation stoichiometry explicitly as in the main 
text,
\begin{gather}
\begin{array}{*{20}c}
   {\mbox{bursty synthesis:}} & {A\xrightarrow{{\nu _1 }}A + b;} & {\nu _1  = \frac{{\gamma}}
{b}\cdot g \left( A \right)},  \\
   {\mbox{linear degradation:}} & {A\xrightarrow{{\nu _2 }}A - 1;} & {\nu _2  =\delta\cdot A,}  \\
 \end{array}
 \label{eq:tslRates_Supp} 
\end{gather}
leading to the propensity vector ${\bm \nu}=\lbrack \frac{\gamma}{b}\cdot g(A), \delta \cdot A\rbrack$ and stoichiometry matrix 
${\bf S}=\lbrack b, -1 \rbrack$. We can easily calculate the coefficient matrices ${\bf \Gamma}$ and ${\bf D}$,
\begin{gather}
{\bf \Gamma}=\lbrack \gamma\cdot g'(A)-\delta\rbrack \quad \quad \quad {\bf D}=\lbrack b\cdot\gamma\cdot g(A)+\delta\cdot 
A\rbrack.
\end{gather}
It is a simple task to then determine the steady-state correlations of the fluctuations,
\begin{gather}
{\bf \Xi}=-\frac{1}{2}\frac{{\bf D}}{{\bf \Gamma}}=-\frac{1}{2}\frac{\lbrack b\cdot\gamma\cdot g(A^{\star})+\delta\cdot 
A^{\star}\rbrack}{\lbrack \gamma\cdot g'(A^{\star})-\delta\rbrack},
\end{gather}
which is positive since the deterministic eigenvalue \mbox{$\lambda=\lbrack \gamma\;g'(A^{\star})-\delta\rbrack <0$} in the stable 
regime where the analysis is carried out. We write the fractional deviation $\eta$ of the steady-state fluctuations in $A$ as,
\begin{gather}
\eta=\frac{\sqrt{\langle A^2 \rangle}}{A^{\star}}=\sqrt {\frac{{\left( {b + 1} \right)}}{{2\left[ {1 - A_0 g' \left( {A^{\star} } 
\right)} \right]}}} \sqrt {\frac{1}
{{A_0\cdot V_{cell}\cdot g \left( {A^{\star} } \right)}}} ,\notag
\end{gather}
where $A^{\star}$ is the steady-state activator concentration and $A_0=\frac{\gamma}{\delta}$ is the fully-activated protein 
concentration and $\omega^{-2}=V_{cell}$ is the cell volume. Provided the {\it HIGH} and {\it LOW} equilibrium points are 
well-separated ($g' \left( {A^{\star} } \right)\approx 0$), we can write,
\begin{gather}
\eta _{LO}  = \sqrt {\frac{{\left( {b + 1} \right)}}
{2}} \sqrt {\frac{f}
{{A_0 \cdot V_{cell} }}}  =  \eta _{HI}\sqrt f,
\end{gather}
where $f$ is the \emph{fold activation}. Not surprisingly, the relative fluctuations around the {\it LOW} state are large since in 
that state, the molecule numbers are small. More importantly for the present discussion, we see that the magnitude of the relative 
fluctuations depends directly upon the burstiness $b$. To determine the effect of the burstiness upon the averaged stability, we 
calculate the stability matrices ${\bf J}^{(0)}$ and ${\bf J}^{(1)}$ (where time has been scaled with respect to the protein 
lifetime: $t\to t\cdot \delta^{-1}$),
\begin{gather}
{\bf J}^{(0)}=\lbrack A_0 \; g'_A(a) - 1\rbrack \quad \quad \omega{\bf J}^{(1)}=\lbrack A_0 \; g''_A(a)\rbrack\omega\;{\bm 
\alpha}(t),\notag
\end{gather}
from which the Laplace transform of the autocorrelation function $\hat{\bf J}_c (s)$ is derived,
\begin{gather}
\omega ^2 {\mathbf{\hat J}}_c \left( s \right) = \omega ^2 \left[ {A_0 g''} \right]^2 \int\limits_0^\infty  {\left\langle {\alpha 
\left( t \right)\alpha \left( 0 \right)} \right\rangle e^{\left[ {A_0 g' - 1} \right]t} e^{ - st} dt} .\notag
\end{gather}
Referring to Eq.~\ref{eq:timeCorrfun_Supp}, the steady-state fluctuations have exponential time-autocorrelation function so that 
the integrand becomes,
\begin{gather}
\omega ^2 {\mathbf{\hat J}}_c \left( s \right) =  - \omega ^2 \left[ {A_0 g''} \right]^2 \frac{{\left( {b + 1} \right)}}
{2}\frac{{A_0 g}}
{{\left[ {A_0 g' - 1} \right]}}\\\notag
\times\int\limits_0^\infty  {e^{\left[ {A_0 g' - 1} \right]t} e^{\left[ {A_0 g' - 1} \right]t} e^{ - st} dt}.\notag
\end{gather}
Evaluating the integral,
\begin{gather}
\omega ^2 {\mathbf{\hat J}}_c \left( s \right) =  - \frac{{\left( {b + 1} \right)}}
{2}\frac{{\omega ^2 }}
{{K_A }}\frac{{A_0^2 g\left[ {A_0 g''} \right]^2 }}
{{\left[ {A_0 g' - 1} \right]}}\frac{K_A}{A_0}\frac{1}
{{s - 2\left[ {A_0 g' - 1} \right]}}.
\end{gather}
From the stability matrices, we are able to calculate the approximation of the effective eigenvalue $\lambda'$ from 
Eq.~\ref{eq:pertEigVals_Supp},
\begin{gather}
\lambda ' = \left[ {A_0 g' - 1} \right] + \frac{{\omega ^2 }}
{{K_A }}\frac{{\left( {b + 1} \right)}}
{2}\frac{K_A}{A_0}\frac{{A_0^4 \left[ {g''} \right]^2 g}}
{{\left[ {A_0 g' - 1} \right]^2 }},
\end{gather}
where we identify $\omega^{-2}=V_{cell}$ as the volume of the cell. Collecting the constants into groups, we write the the 
effective eigenvalue $\lambda '(A^{\star})$ as,
\begin{gather}
\lambda ' = \lambda  + \frac{1}{V_{cell}} \lambda _{corr}  = \lambda \left\{ {1 - \Delta_b \cdot h\left( {\frac{{A_0 }}
{{K_A }},g\left( {A^{\star}} \right)} \right)} \right\},
\label{eq:effEigenTsl}
\end{gather}
where $\Delta_b=\frac{(b+1)}{2}\frac{1}{K_A\cdot V_{cell}}$ is the discrete change in reactant molecule numbers, scaled with 
respect to the number of activators required to initiate activation ($K_A\cdot V_{cell}$), representing the relative change in 
protein numbers incurred by the stochastic reaction events. (In a sense, $K_A$ represents the characteristic concentration of the 
activator: for activator concentrations far less than $K_A$, there is no activation and for concentrations far above $K_A$, the 
promoter is fully activated.) The second term in Eq.~\ref{eq:effEigenTsl}, \mbox{$h\left( {\frac{{A_0 }}
{{K_A }},g\left( {A^{\star}} \right)} \right) = \frac{K_A}{A_0}\frac{{A_0^4 \left( {g''} \right)^2 g}}
{{\left| \lambda  \right|^3 }}$} contains the details of the regulatory mechanism~\cite{BintuMODELS} and depends strongly upon the 
stability of the deterministic system through $\lambda$. It is the interplay between the fluctuations (through $\Delta_b$) and the 
macroscopic stability of the steady-state (through $h$) that ultimately decides the averaged stability of the stochastic system. 

\subsubsection{Accuracy of ESA} 

\begin{figure}
\begin{center}
\epsfig{file=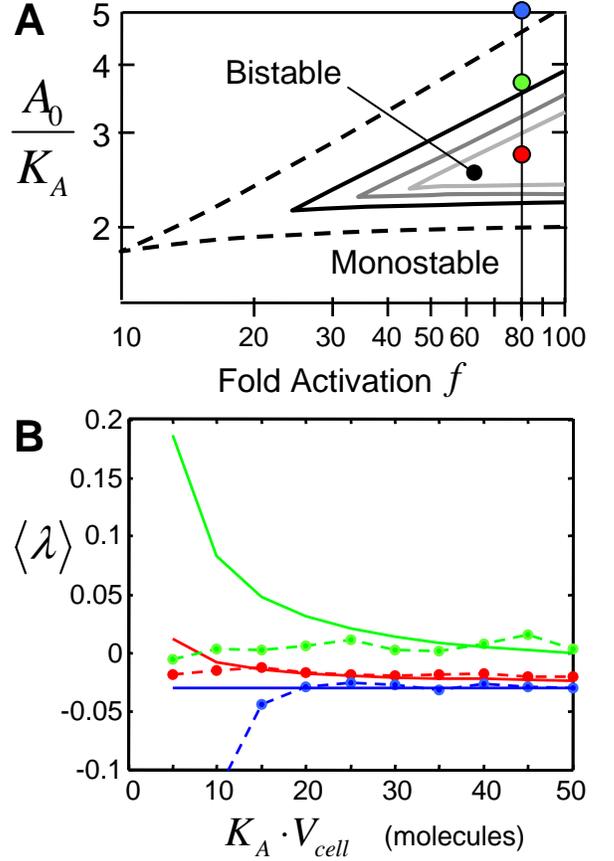,width=0.9\linewidth}
\caption{Accuracy of the effective stability approximation (ESA) as a function of the number of molecules. (A) Focusing upon three 
points in the parameter space of the autoactivator model (see Figure 2a in the main text), it is possible to compare the ESA with 
the results of numerical simulation. (B) The short-time Lyapunov exponent of an ensemble average of the perturbation modes about 
the LOW state (dashed lines) approach those values of $\lambda'$ predicted according to Eq. 25 (solid lines) for systems with 
increasing values of $K_A\cdot V_{cell}$, which specifies the order of molecule numbers to turn on/off the gene. Here, the 
burstiness of protein synthesis is held constant at $b=9$, and each data point is computed from a sample of $10^5$ trajectories -- 
colors of the curves correspond to the filled circles in panel A.}
\end{center}
\end{figure}

To compute the accuracy of the effective stability approximation as a function of the molecule numbers for the translational 
autoactivator model, the corrected eigenvalue $\lambda'$ computed above (Eq.~\ref{eq:effEigenTsl}) is compared to the short-time 
Lyapunov exponent of the ensemble-averaged perturbation modes computed by stochastic simulation~\cite{Gillespie}.

For a system slightly perturbed from the steady-state $x_s$, the short-time Lyapunov exponent $\langle \lambda \rangle$ is defined 
as,
\begin{gather*}
\mathop {\lim }\limits_{t \to 0} \ln \left| {\left\langle {x_p \left( t \right)} \right\rangle  - x_s } \right| = \mbox{const.} + 
\left\langle \lambda  \right\rangle  \cdot t.
\end{gather*}
A numerical calculation of $\langle \lambda \rangle$ is obtained by taking the ensemble average (over an ensemble of $10^5$ 
members) of $x_p(t)$ determined by stochastic simulation. The slope of the natural-log difference between the numerically 
generated perturbation mode and the steady state, $\ln |\langle x_p(t)\rangle -x_s|$, is fit by linear regression over a time span 
corresponding to the protein lifetime ({\it i.e.} $\delta^{-1}=30$ minutes). To compare the stochastic simulation with the ESA, we 
focus upon three points in the parameter space of the autoactivator (Figure 1a, filled circles) -- one point well inside the 
bistable regime ($\frac{A_0}{K_A}=2.5,f=80$; red), one near the boundary predicted by the ESA ($\frac{A_0}{K_A}=3.5,f=80$; green), 
and one well inside the monostable regime ($\frac{A_0}{K_A}=5,f=80$; blue). Figure 1b compares the resulting Lyapunov exponent 
$\langle \lambda \rangle$ (dashed lines) with the ESA prediction $\lambda'$ (solid lines), where the line colors correspond to the 
colors of the filled circles in Figure 1a. Here, the burstiness in protein synthesis is held constant at $b=9$, and the 
characteristic number of molecules in the system, $K_A\cdot V_{cell}$, is increased from 5 to 50. (In the main text, $K_A\cdot 
V_{cell}=25$ so that a burstiness of $b=9$ gives a discreteness parameter of $\Delta_{b_A}=\frac{(b+1)}{2}\frac{1}{K_A\cdot 
V_{cell}}=0.2$.) As the number of molecules in the system is increased, the ESA and the numerical simulation results converge.  
The figure shows the effective stability of the transcriptional autoactivator model is well-characterized by the ESA for systems 
with $K_A\cdot V_{cell}\gtrsim 20$.

\subsubsection{Translational activation}

To model the translational activity, we redefine the transcription rate to be constant $\frac{\gamma}{b}$, where $b$ is the 
maximum burst size at full activation, and allow the activator to control the translation rate through the \emph{stoichiometery}. 
We write the synthesis and degradation reactions -- in analogy with Eq.~\ref{eq:tslRates_Supp} above -- as, 
\begin{gather}
\begin{array}{*{20}c}
    {A\xrightarrow{{\nu _1 }}A + b  \cdot g\left( A \right);\quad} & {\nu _1  = \frac{{\gamma}}
{{b }}},  \\
   {A\xrightarrow{{\nu _2 }}A - 1;\quad} & {\nu _2  = \delta\cdot A},  \\
 \end{array} 
\end{gather}
where the translational activation affects the stoichiometry through the synthesis step-size $b \cdot g(A)$. Notice that the 
deterministic equation \mbox{$\frac{dA}{dt}={\bf S}\cdot{\bm \nu}=A_0\;g(A)-A$} is \emph{identical} to the deterministic equation 
for the transcriptional autoactivator in the previous section. Nonetheless, the change in synthesis stoichiometry from $b \mapsto 
b  \cdot g\left( A \right)$ has a noticeable effect on the resulting stability. As above, we calculate the effective eigenvalue,
\begin{gather}
\lambda ' = \lambda \left\{ {1 - \frac{{\left( {b \cdot g\left( A^{\star} \right) + 1} \right)}}
{2}\frac{1}
{{V_{cell}\cdot K_A }} \cdot h\left( {\frac{{A_0 }}
{{K_A }},g\left( {A^{\star}} \right)} \right)} \right\},\notag
\end{gather}
where $h(\;\cdot\;)$ is as in Eq.~\ref{eq:effEigenTsl}. The difference from the transcriptional case is that the burst-size itself 
is attenuated in the {\it LOW} state, and the discreteness parameter approaches the minimal value $\Delta_b \to 1 / (2 
V_{cell}\cdot K_A)$, thereby increasing the residence time in the {\it LOW} state.

\subsection{Genetic oscillator}

The parameters of Vilar {\it et al.}~\cite{Vilar} correspond to the reduced model parameters:
\begin{gather}
\gamma_{A}=25\;nM\;h^{-1}, K_A=0.5\;nM, f_A=10, \\\notag
\gamma_{R}=5\;nM\;h^{-1}, K_R=1\;nM, f_R^{-1}=0,\\\notag
 \kappa_C=2\times 10^{2}\;nM^{-1}\;h^{-1}, \mbox{and } \delta_A=1\;h^{-1},
\end{gather}
where, for simplicity, we make the approximation that 1 molecule / $1 \mu m^3 \approx 1\; nM$ and set $V_{cell}=100 \mu m^3$. 
Furthermore, the mRNA degradation and translation rates in the original model give an activator burst size of $b_A=5$ and a 
repressor burst size of $b_R=10$. 

\subsubsection{Details of the stochastic model}

The reduced model (Eq. 6 in the main text) is composed of six elementary reactions:
\begin{gather}
\begin{array}{*{20}c}
   {A \to A + b_A} & {\nu_1=\frac{{\gamma _A }}{{b_A }} \cdot g\left( {\frac{A}{{K_A }},f_A } \right)}  \\
   {A\to A - 1} & {\nu_2=\delta _A  \cdot A}  \\
   {\left({A,R,C}\right)\to \left({A-1,R-1,C+1}\right)} & {\nu_3=\kappa _C  \cdot A \cdot R}  \\
   {R \to R + b_R} & {\nu_4=\frac{{\gamma _R }}{{b_R }} \cdot g\left( {\frac{A}{{K_R }},f_R } \right)}  \\
   {R \to R -1} & {\nu_5=\delta _R  \cdot R}  \\
   {\left({R,C}\right)\to \left({R+1,C-1}\right)} & {\nu_6=\delta _A  \cdot C}  \\
\end{array}\notag
\end{gather}
The stoichiometry matrix ${\bf S}$ and the propensity vector ${\bm \nu}$ are then written as,
\begin{gather}
{\bf{S}} = \left[ {\begin{array}{*{20}c}
   {b_A } & { - 1} & { - 1} & 0 & 0 & 0  \\
   0 & 0 & { - 1} & {b_R } & { - 1} & 1  \\
   0 & 0 & 1 & 0 & 0 & { - 1}  \\
\end{array}} \right],\\\notag
 \bm{\nu}  = \left[ {\begin{array}{*{20}c}
   {\frac{{\gamma _A }}{{b_A }} \cdot g\left( {\frac{A}{{K_A }},f_A } \right)}  \\
   {\delta _A  \cdot A}  \\
   {\kappa _C  \cdot A \cdot R}  \\
   {\frac{{\gamma _R }}{{b_R }} \cdot g\left( {\frac{A}{{K_R }},f_R } \right)}  \\
   {\delta _R  \cdot R}  \\
   {\delta _A  \cdot C}  \\
\end{array}} \right].
\end{gather}
Identification of dimensionless parameters in the deterministic model comes from considering the rate equations, 
\begin{gather}
\frac{d}{{dt}}\left[ {\begin{array}{*{20}c}
   A  \\
   R  \\
   C  \\
\end{array}} \right] = {\bf{S}} \cdot \bm{\nu}  = \\\notag
\left[ {\begin{array}{*{20}c}
   {\gamma _A  \cdot g\left( {\frac{A}{{K_A }},f_A } \right) - \delta _A  \cdot A - \kappa _C  \cdot A \cdot R}  \\
   {\gamma _R  \cdot g\left( {\frac{A}{{K_R }},f_R } \right) - \delta _R  \cdot R - \kappa _C  \cdot A \cdot R + \delta _A  \cdot 
C}  \\
   {\kappa _C  \cdot A \cdot R - \delta _A  \cdot C}  \\
\end{array}} \right].
\end{gather}
In what follows, it will be convenient to call $\gamma=\frac{\gamma_R}{\gamma_A}$ and $A_0=\frac{\gamma_A}{\delta_A}$. Scaling the 
concentrations with respect to the characteristic concentration $A_0$ ({\it i.e.} $A=A'\cdot A_0$, {\it etc.}) and time with 
respect to the activator lifetime, $t=t'\cdot \delta_A$, the rate equations become,
\begin{gather}
\frac{d}{{dt'}}\left[ {\begin{array}{*{20}c}
   A'  \\
   R'  \\
   C'  \\
\end{array}} \right] = \\
\left[ {\begin{array}{*{20}c}
   {g\left( {A'\frac{{A_0 }}{{K_A }},f_A } \right) - A' - \left[ {\frac{{\kappa _C  \cdot A_0 }}{{\delta _A }}} \right] \cdot A' 
\cdot R'}  \\
   {\gamma \cdot g\left( {A'\frac{{A_0 }}{{K_R }},f_R } \right) - \left[ {\frac{{\delta _R }}{{\delta _A }}} \right] \cdot R' - 
\left[ {\frac{{\kappa _C \cdot A_0 }}{{\delta _A }}} \right] \cdot A' \cdot R' + C'}  \\
   {\left[ {\frac{{\kappa _C  \cdot A_0 }}{{\delta _A }}} \right] \cdot A' \cdot R' - C'}  \\
\end{array}} \right].\notag
\end{gather}
The two additional dimensionless constants are the scaled rate of dimerization $\kappa  = \frac{{\kappa _C  \cdot A_0 }}{{\delta 
_A }}$ and the ratio of the repressor and activator degradation rates $\epsilon  = \frac{{\delta _R }}{{\delta _A }}$. Henceforth, 
the primes denoting the dimensionless quantities will be dropped.

Since the variance in the fluctuations is found from the auxiliary matrices ${\bf \Gamma}$ and ${\bf D}$ ({\it cf.} Eq. 24), and 
${\bf \Gamma}$ is the Jacobian of the deterministic system, the dimensionless stochastic parameters are most easily found by 
considering ${\bf D} ={\bf S}\cdot \mbox{diag}\left[{ \bm{\nu} }\right] \cdot {\bf S}^T$,
\begin{widetext}
${\bf D}=\left[{{\begin{array}{*{20}c}
   {b_A  \cdot \gamma _A  \cdot g_A  + \delta _A  \cdot A + \gamma _C  \cdot A \cdot C} & {\gamma _C  \cdot A \cdot C} & { - 
\gamma _C  \cdot A \cdot C}  \\
   {\gamma _C  \cdot A \cdot C} & {b_R  \cdot \gamma _R  \cdot g_R  + \delta _R  \cdot R + \gamma _C  \cdot A \cdot C + \delta _A  
\cdot C} & { - \gamma _C  \cdot A \cdot C - \delta _A  \cdot C}  \\
   { - \gamma _C  \cdot A \cdot C} & { - \gamma _C  \cdot A \cdot C - \delta _A  \cdot C} & {\gamma _C  \cdot A \cdot C + \delta 
_A  \cdot C}  \\
\end{array}}}\right],$
\end{widetext}
where $g_i  \equiv g\left( {\frac{A}{{K_i }},f_i } \right)$. As above, we scale the concentrations with respect to $A_0$ and 
divide through by $\delta_A$. Evaluating ${\bf D}$ at the steady-state $\left( {A^{\star} ,R^{\star} ,C^{\star} } \right)$, where 
$\frac{{dA}}{{dt}} = \frac{{dR}}{{dt}} = \frac{{dC}}{{dt}} = 0$, provides the additional simplifications derived from the rate 
equations above, written in dimensionless form,
\begin{eqnarray}
 g_A & = & A^{\star}  + \kappa  \cdot A^{\star}  \cdot R^{\star} , \\ \notag
 \gamma\cdot g_R  + C^{\star} & = & \epsilon  \cdot R^{\star}  + \kappa  \cdot A^{\star}  \cdot R^{\star},  \\ \notag
 C^{\star} & = & \kappa  \cdot A^{\star}  \cdot R^{\star} .
 \end{eqnarray}
Hence, the matrix ${\bf D}$ is written in terms of reactant \emph{numbers} as,
\begin{gather}
\frac{{\bf{D}}}{{\gamma  \cdot A_0 }} = \left[ {\begin{array}{*{20}c}
   {2\left[ {\frac{{\left( {b_A  + 1} \right)}}{2}} \right]\frac{{g_A }}{\gamma }} & {C^{\star} } & { - C^{\star} }  \\
   {C^{\star} } & {2\left[ {\frac{{\left( {b_R  + 1} \right)}}{2}} \right]g_R  + 2C^{\star} } & { - 2C^{\star} }  \\
   { - C^{\star} } & { - 2C^{\star} } & {2C^{\star} }  \\
\end{array}} \right].
\end{gather}
Comparing each diagonal element with the characteristic mean reactant number of that species $(N_A\sim K_A\;V_{cell}$, $N_R\sim 
K_R\;V_{cell})$, and ignoring parameters coming from the deterministic model ($g_A, g_R, \mbox{and } \gamma$), we have three 
additional constants - the discreteness in the activator number $\Delta _{b_A }  = \frac{{\left( {b_A  + 1} 
\right)}}{2}\frac{1}{{K_A  \cdot V_{cell} }}$, the discreteness in the repressor number $\Delta _{b_R }  = \frac{{\left( {b_R  + 
1} \right)}}{2}\frac{1}{{K_R  \cdot V_{cell} }}$ and the extent of dimerization $\frac{{C^{\star} }}{{K_R  \cdot V_{cell} }}$. In 
the main text, we focus upon the effect of varying the deterministic parameter $\epsilon$ and the stochastic parameter 
$\Delta_{b_A}$.

\section{Algorithmic Implementation of the the Effective Stability Approximation}

The corrections to the deterministic eigenvalues are computed by solving the resolvent equation for the the effective eigenvalues 
$\lambda'$,
\begin{gather}
\mbox{det}\lbrack {\lambda' \cdot {\bf I} - {\bf J}^{(0)} - \frac{1}{V_{cell}}\hat {\bf J}_c (\lambda')} \rbrack,
\end{gather}
(Eq. 12 in the main text). In this section, we provide a step-by-step algorithm to form the matrices ${\bf J}^{(0)}$ and $\hat 
{\bf J}_c (\lambda')$ from the deterministic reaction rates. In the following, the deterministic state vector is denoted by {\bf 
x} and ${\bm \alpha}$ denotes the fluctuations in each of the components of ${\bf x}$ ({\it c.f.} Section I-C above). The first 
three steps of the algorithm come from the paper by Elf and Ehrenberg~\cite{Elf}.

\begin{enumerate}
\item Write the various reactions in terms of their \emph{propensity} and \emph{stoichiometry}. The deterministic reaction rates 
are formed by the product ${\bf S}\cdot{\bm \nu}$ ({\it cf.} Eqs. 31 and 41 above).

\item From ${\bf S}$ and ${\bm \nu}$, construct the matrices ${\bf \Gamma}$ and ${\bf D}$,
\begin{gather}
{\bf \Gamma}_{ij}({\bf x})=\frac{\partial\lbrack{{\bf S}\cdot {\bm \nu}}\rbrack_{i}}{\partial x_j} \quad \quad {\bf D}({\bf 
x})={\bf S}\cdot \mbox{diag}[{\bm \nu}]\cdot {\bf S}^T.
\end{gather}

\item Compute the steady-state covariance in the fluctuations $\bm{\alpha}$ by solving the fluctuation-dissipation relation for 
each of the entries in the symmetric covariance matrix ${\bf \Xi}$ (where $\Xi_{ij}=\Xi_{ji}=\langle{\alpha_i\;\alpha_j}\rangle$),
\begin{gather}
{\bf \Gamma}({\bf x}_s)\cdot {\bf \Xi} + {\bf \Xi}\cdot {\bf \Gamma}^T({\bf x}_s) +{\bf D}({\bf x}_s)={\bf 0}.
\end{gather} 
The steady-states ${\bf x}_s$ are calculated from the deterministic reaction rates by solving the algebraic equations 
$(\lbrack{{\bf S}\cdot {\bm \nu}}\rbrack_{{\bf x}={\bf x}_s})={\bf 0}$.

Evaluated at the steady-state, the fluctuation-dissipation relation is simply a $\frac{1}{2}d(d+1)$ system of linear equations 
that determine the symmetric entries of ${\bf \Xi}$ (where $d$ is the dimension of the system). For more details regarding the 
general solution of the fluctuation-dissipation relation, see~\cite{Tomioka}. 

\item Compute the matrices ${\bf J}^{(0)}$ and ${\bf J}^{(1)}(t)$,
\begin{gather}
{\bf J}^{(0)}={\bf \Gamma}({\bf x}_s)\quad\quad{\bf J}^{(1)}(t)=\frac{\partial {\bf \Gamma}({\bf x}_s+\omega\;{\bm 
\alpha}(t))}{\partial \omega}|_{\omega=0}.
\end{gather}

\item Calculate the matrix ${\bf J}_c(t)$,
\begin{gather}
{\bf J}_c(t)=\langle{{\bf J}^{(1)}(t)\cdot \mbox{exp}\lbrack{{\bf J}^{(0)}\;t}\rbrack\cdot {\bf J}^{(1)}(0)}\rangle,
\end{gather}
where $\mbox{exp}\lbrack{{\bf J}^{(0)}t}\rbrack$ is the matrix exponential of ${\bf J}^{(0)}$. The matrix ${\bf J}_c(t)$ will be 
composed of linear combinations of the autocorrelation functions $\langle{\alpha_i(t)\;\alpha_j(0)}\rangle$. Replace each of these 
by the ${(i,j)}^{th}$ element of the matrix $\mbox{exp}\lbrack{{\bf J}^{(0)}\;t}\rbrack\cdot {\bf \Xi}$,
\begin{gather}
\langle{\alpha_i(t)\;\alpha_j(0)}\rangle=\lbrack{\;\mbox{exp}[{\bf J}^{(0)}\;t] \cdot {\bf \Xi}\;}\rbrack _{ij},
\end{gather}
({\it cf.} Eq. 25 above).

\item The correction matrix ${\bf J}_c(t)$ is composed of exponential terms of the form $e^{at}$, facilitating the computation of 
the Laplace transform $\hat{\bf J}_c(\lambda')$. Simply replace each term $e^{at}$ with $(\lambda'-a)^{-1}$,
\begin{gather}
\hat{\bf J}_c(\lambda')={\bf J}_c(t)|_{e^{at}\to(\lambda'-a)^{-1}}.
\end{gather}

\item Solve the resolvent equation for $\lambda'$,
\begin{gather}
\mbox{det}\lbrack {\lambda' \cdot {\bf I} - {\bf J}^{(0)} - \frac{1}{V_{cell}}\hat {\bf J}_c (\lambda')} \rbrack.
\end{gather}

\end{enumerate}

The algorithm described above is easily implemented in symbolic mathematics packages. A version coded in {\it Mathematica} is 
available from the authors upon request.


\bibliographystyle{apsrev}
\bibliography{pnasbib}

\end{document}